\documentclass[draft,sw]{agutex}

\usepackage{color}
\usepackage[dvips]{graphicx}
\setkeys{Gin}{draft=false}
\authorrunninghead{SCHMIDT AND CAIRNS}

\titlerunninghead{SPACE WEATHER RELEVANT FEATURES}

\authoraddr{Corresponding author: J. M. Schmidt, School of Physics, University of Sydney, NSW 2006, Australia. (joachim.schmidt@sydney.edu.au)}

\begin{document}

%
%

\title{{\color{black} Hit or Miss, Arrival Time, and $B_z$ Orientation Predictions} of BATS-R-US CME Simulations {\color{black} at 1 AU}}
%
%

%
%









\authors{J. M. Schmidt\altaffilmark{1} and Iver H. Cairns\altaffilmark{1}}

\altaffiltext{1}{School of Physics, University of Sydney, NSW 2006, Australia.}

%
%


\begin{abstract}
{\color{black} Using a refined setup process,} we simulated the propagation of six observed Coronal Mass Ejections (CMEs) with the 2012 Block-Adaptive-Tree-Solarwind-Roe-Upwind-Scheme (BATS-R-US) code from {\color{black} the} Sun to {\color{black} the} Earth or STEREO A and {\color{black} compared the outputs with observations.} {\color{black} A linear relation between the average CME speed below 6 solar radii and the flux rope current is demonstrated and used to tune the simulations.} The simulations correctly predict if and when an observable CME shock reaches one astronomical unit (AU). {\color{black} The arrival} time predictions of the CME {\color{black} shocks} at 1 AU have an accuracy of 0.9 $\pm$ 1.9 hours. The simulated initial CME speeds and {\color{black} average} accelerations are close to the model and data of \citet{Gopalswamy00}. The approach shows promise for predicting the sense of the predominant shock-associated change in the magnetic field component $B_z$. {\color{black} However, the magnetic fields and plasma conditions in the solar wind and CME are not predicted well quantitatively.} 
\end{abstract}

%
%

%

\begin{article}

%
%

\section{Introduction}

Coronal Mass Ejections (CMEs) hitting Earth's magnetosphere can trigger space weather events and major geomagnetic storms that lead to major disruptions in ground- and space-based electrical systems and devices \citep[see, e.g.,][]{Schrijver15}. The mitigation of impacts requires forecasting if and when a CME will hit the Earth, and what the plasma and magnetic field variations will be. A widely used tool for such forecasting is the ENLIL code developed at the Space Weather Prediction Center (SWPC) of the National Oceanic and Atmospheric Administration (NOAA), USA. This code numerically simulates a CME launched at the Sun and propagating through the interplanetary space to one astronomical unit (AU) and beyond, including planets and spacecraft of interest. {\color{black} The} Sun - CME system is {\color{black} usually} described {\color{black} with} Magneto HydroDynamics (MHD). However, in the ENLIL code the magnetic fields of the CME are either neglected or included as a teardrop shaped spheromak that has radial instead of azimuthal magnetic fields at the front of the CME \citep[see, e.g.,][]{Odstrcil15,Odstrcil18}. This makes the numerical calculations fast, but leads to significant differences from the observations. Specifically, the predicted arrival time of the CME at Earth differs from that observed by 10 $\pm$ 1 hours {\color{black} on average} \citep[see, e.g.,][]{Wold18,Verbeke18,Riley18}. Even so, the ENLIL code is the default approach internationally for simulating the motion of CMEs. 

Another sophisticated simulation code for a Sun - CME system that includes full MHD is the Block-Adaptive-Tree-Solarwind-Roe-Upwind-Scheme (BATS-R-US) code and its subsequent Space Weather Modeling Framework (SWMF) extension developed at the University of Michigan \citep[]{Powell99,Roussev03,Roussev04,Cohen07,Cohen08,Toth12}.
{\color{black} Considerable efforts have been made to simulate realistic CMEs with this code \citep[see, e.g.,][]{Manchester08, Manchester12}.}   

{\color{black} Our major focus in this paper is on operational users who need to be able to accurately predict whether and when a CME will reach Earth and only then address the detailed properties of the CME (e.g., $B_z$).}
For this purpose we {\color{black} developed a refined event-specific simulation setup approach for the 2012 BATS-R-US code and} simulated six CMEs that were launched towards the Earth or STEREO A. The simulation setup is such that the time needed to {\color{black} set up and} carry out such a simulation is smaller {\color{black} ($< \approx 1$ day)} than the actual observed travel time of the CME to 1 AU, typically 2-3 days. This includes the event-specific setup of the simulation, the running of the code, and the analysis and interpretation of the results. This setup and analysis is {\color{black} refined from} our earlier work {\color{black} simulating type II radio bursts and CMEs} \citep[][]{Schmidt13,Schmidt14a,Cairns15,Schmidt16a,Schmidt16b}. Crucial simulation outputs are whether or not the CME reaches Earth, the arrival time of a CME at 1 AU, and the north-south component $B_z$ with respect to the ecliptic plane before, during, and after the CME's arrival at 1 AU. Reliable prediction of the CME's motion and arrival time at 1 AU enables the forecasting of if and when the CME can drive a space weather event at Earth. The correct prediction of the orientation of $B_z$ during the event enables the forecast of whether an incident CME should trigger a space weather event or not. If $B_z$ is oriented {\color{black} southward}, then magnetic reconnection may occur at Earth's magnetopause and magnetotail, triggering geomagnetic storms and other disruptions in the magnetosphere. 

In section 2 of this paper we describe a refined setup to simulate the propagation of CMEs with the 2012 version of the BATS-R-US code from the Sun to 1 AU. We demonstrate that a linear relation exists between the average CME speed below 6 $R_s$ and the flux rope current for each event-specific setup and use this to tune the simulations to the observed CME. We correctly predict if and when an observable CME shock reaches 1 AU (section 3). The approach also shows promise in predicting the sense of the predominant shock-associated change $B_z$, but, in common with many other codes and reports does not adequately model the background solar wind or CMEs.

\section{Simulation setup}

In the BATS-R-US code the solar coronal magnetic fields are reconstructed with a truncated {\color{black} spherical harmonic} series expansion of the magnetic fields measured with the Wilcox Solar Observatory [see, e.g., wso.stanford.edu]. The analytic coefficients of this series are listed in a file for each Carrington rotation of the Sun and stored in an archive at the Wilcox Solar Observatory. The BATS-R-US code reads the specific coefficient file during a simulation run {\color{black} of} a specific CME.

The reconstruction of the solar wind magnetic field, speed, and density to 1 AU and beyond in the 2012 version of {\color{black} BATS-R-US} is based on the model of \citet{Cohen07,Cohen08}. {\color{black} This} uses (1) a modified Wang, Sheeley and Arge model \citep[]{Arge00} in order to obtain {\color{black} initial} estimates for magnetic field, speed, and density {\color{black} profiles}, and (2) Bernoulli integrations along the magnetic field lines in order to estimate the ratio of specific heats for the plasma and so improved estimates for the solar wind speeds and densities.

The CME is then introduced into the reconstructed solar wind as an analytic {\color{black} \citet{Titov99}} flux rope, which is launched by cutting the magnetic field lines that anchor the flux rope to the {\color{black} Sun}. This analytic flux rope is dimensioned with parameters derived from observations of the initial CME. We determined these parameters {\color{black} with a refined version of} the CME Analysis Tool (CAT) in SolarSoft \citep[see, e.g.,][]{Millward13}. This refined tool reads STEREO A/COR2, STEREO B/COR2, SOHO/LASCO C2, and/or SOHO/LASCO C3 coronagraph images of the CME and adjusts a cone interactively to fit the CME. {\color{black} Each CME fit assumes an aspect ratio of 0.8 for the flux rope cross section.} 

By applying the cone fit to successive images of a specific CME, a height-time diagram of the erupting CME is created with CAT. The averaged slope of that height-time diagram below 6 solar radii ($R_s$) is a measure of the CME's initial velocity. In the simulation, the CME's outflow velocity is a function of a solar subsurface current that generates the CME flux rope and the strength of the reconstructed surrounding solar magnetic fields. For the setup of {\color{black} each} CME we varied the subsurface current and determined the resulting {\color{black} CME speed} as the slope of {\color{black} the CME's} height-time {\color{black} diagram} below 6 $R_s$. 

\begin{figure}[!ht]
\includegraphics[scale=1.0]{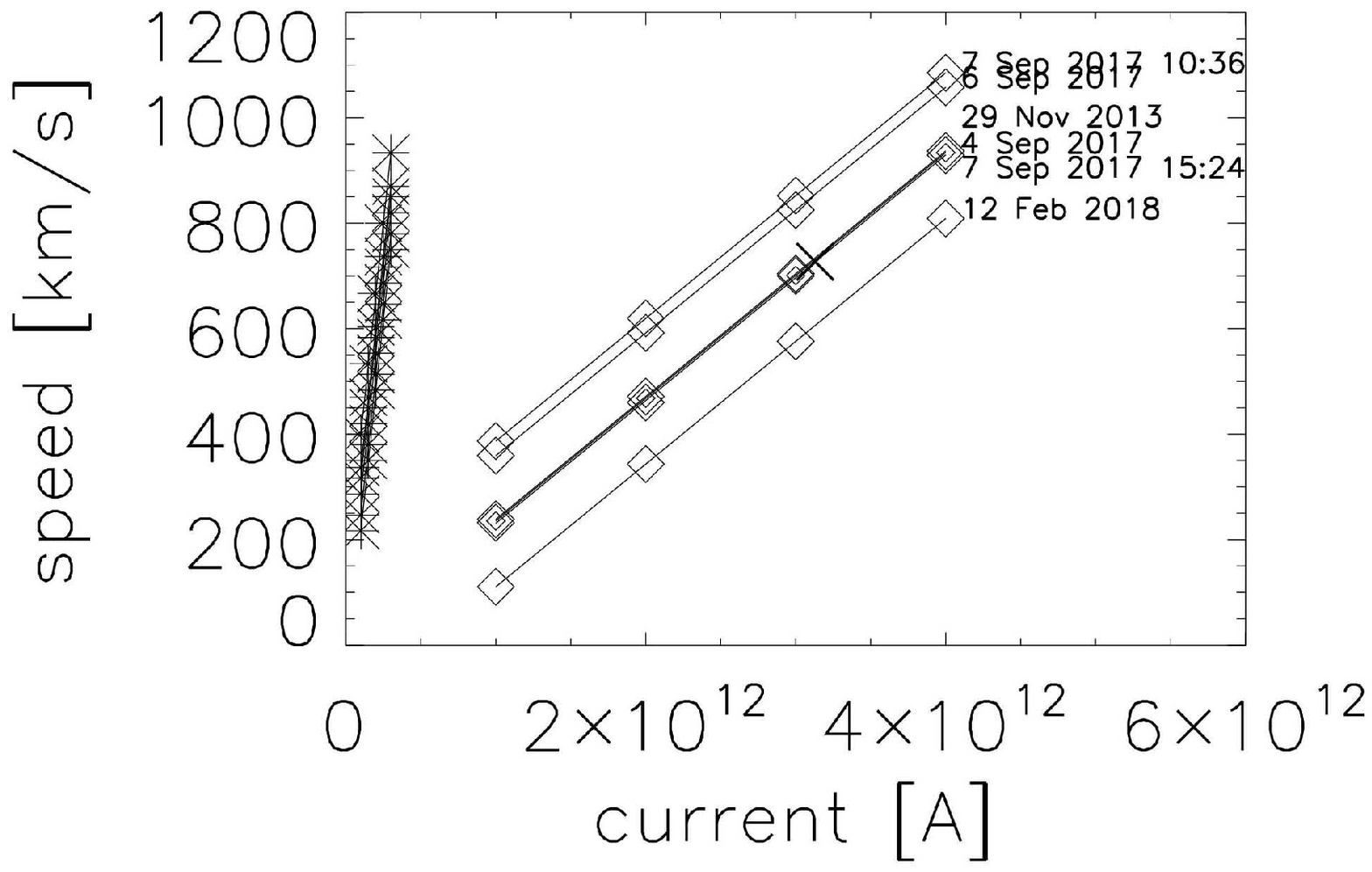}
\caption{\label{Figure_1} Simulated CME outflow speeds (diamonds) as a function of the solar subsurface current for the six CME simulations {\color{black} with level 5 grid refinement}. The relationship is {\color{black} closely} linear. {\color{black} The stars are for runs with level 2 grid refinement, and 'X' denotes the fine-tuned value for the 29 Nov 2013 event.}}
\end{figure}

{\color{black} A crucial result, illustrated in Fig.~\ref{Figure_1} for the six simulated events, is a closely} linear relationship between the subsurface current and the {\color{black} average} CME speed {\color{black} below 6 $R_s$}. The lines that connect the diamonds belong to simulations with an initial level 5 grid resolution, and the lines that connect the stars belong to simulations with an initial level 2 grid resolution.
Note that both the level 2 and level 5 simulations show linear relations between current and CME velocity, differing substantially in slope between level 2 and level 5 runs but being very similar between events with the same initial grid refinement level. 
The linear relationship for each specific event and a given grid refinement level {\color{black} was} then used to tune the subsurface current required to {\color{black} match} the observed CME speed, which was evaluated using the CAT tool. No other tuning was performed.
Note that only 3 runs are thus required to simulate a given event; 2 to define the linear relation and 1 for the fine-tuned run. 

{\color{black} Another important refinement was to} increase the number of grid cells in the simulation to about 16 million cells from the default of about 32,000 cells, corresponding to  {\color{black} increasing the initial grid refinement level from two to five and an} average spatial grid resolution of about 1 $R_s$ (or 1 hour convection time) at 1AU. This reduces the {\color{black} numerical} dispersion of the code and allows simulation of sharp CME-driven shocks that define precise arrival times at 1 AU. 

\begin{figure}[!ht]
\begin{center}
\includegraphics[scale=0.09]{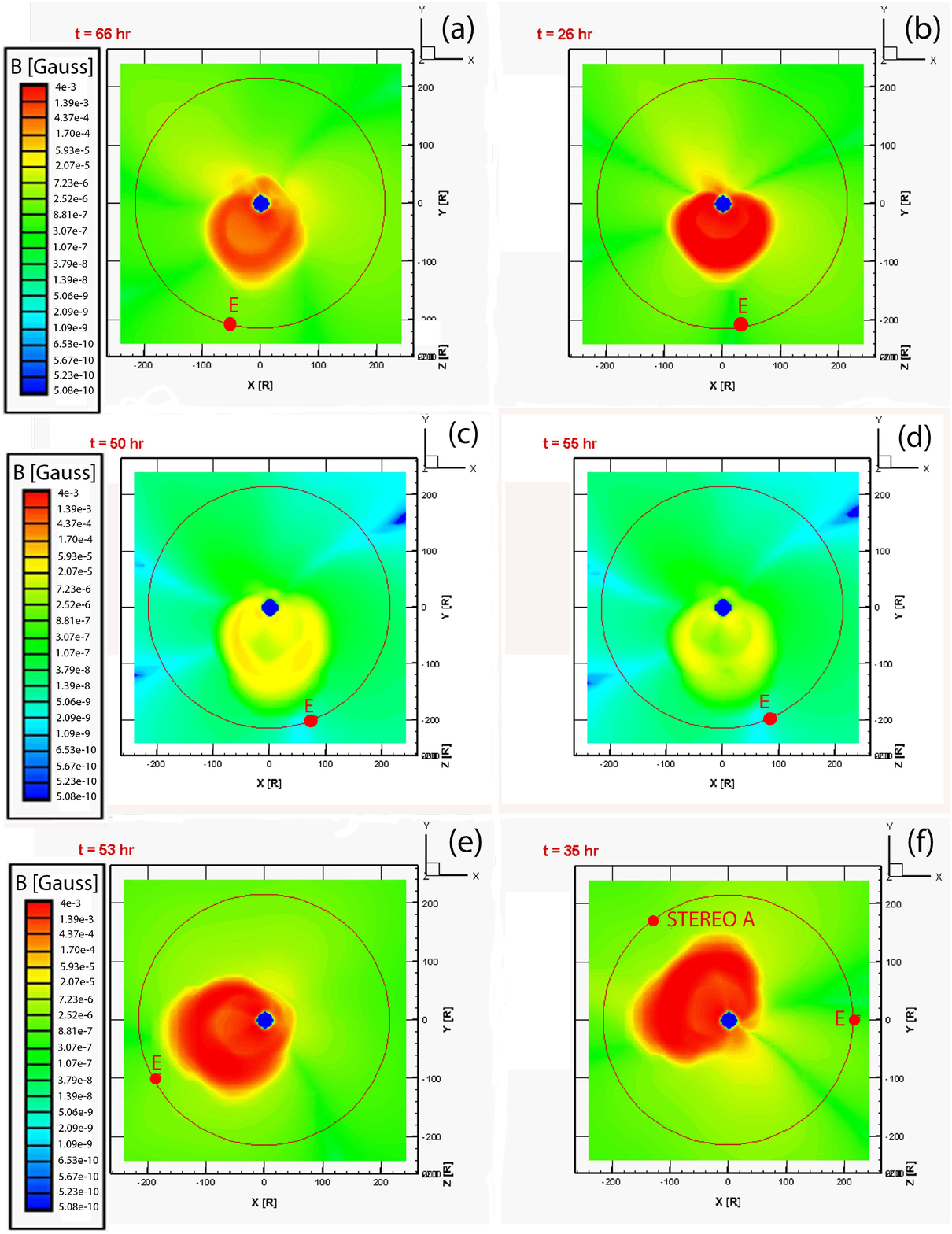}
\end{center}
\caption{\label{Figure_2} Simulated magnetic fields in the ecliptic plane for the CME events of (a) 4 Sep 2017 19:12, (b) 6 Sep 2017 12:30, (c) 7 Sep 2017 10:36, (d) 7 Sep 2017 15:24, (e) 12 Feb 2018 2:00, and (f) 29 Nov 2013 20:00, propagating toward the Earth (a)-(e) and STEREO A (f).
The {\color{black} central} blue dot in each panel is a sphere of 20 $R_s$ surrounding the Sun. The red circle is the orbit of the Earth with a red dot  {\color{black} labelled} E {\color{black} marking} Earth. In Fig.~\ref{Figure_2}(f) {\color{black} the} red dot denotes the position of STEREO A.}
\end{figure}

\begin{figure}[!ht]
\includegraphics[scale=0.18]{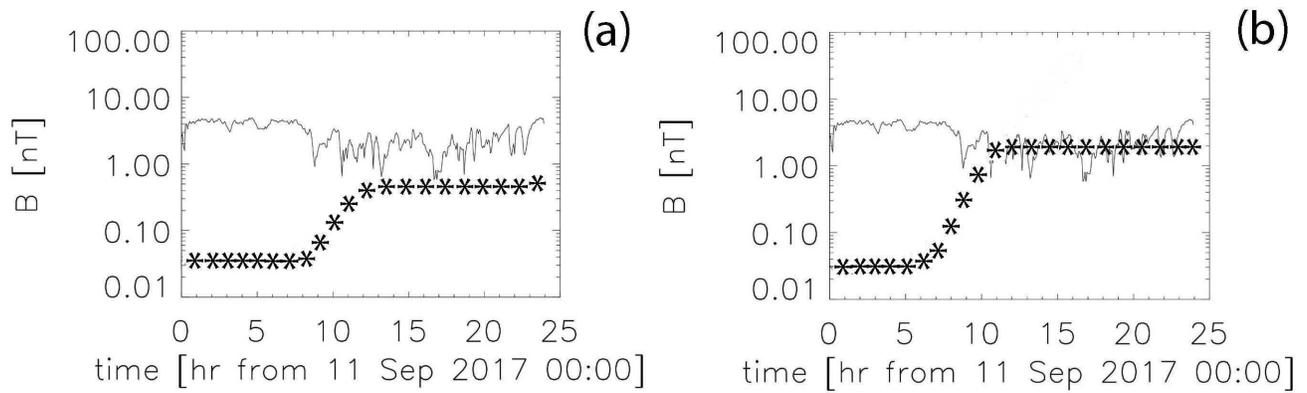}
\caption{\label{Figure_3} Simulated magnetic field (stars) and ACE data (solid lines) at Earth's orbit for the (a) 7 Sep 2017 10:36 and (b) 7 Sep 2017 15:24 CME events. The simulated field remains below or near the observed background, which shows no signature of a shock arrival.}
\end{figure}

\begin{figure}[!ht]
\includegraphics[scale=1.0]{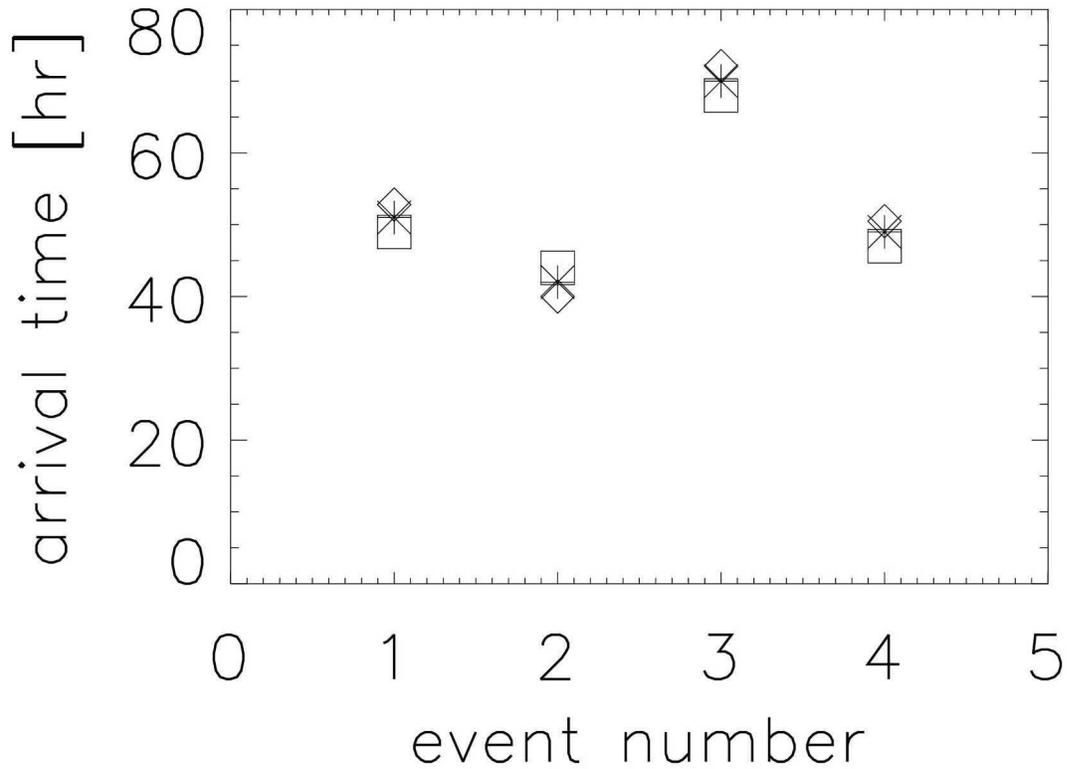}
\caption{\label{Figure_4} Observed (simulated) arrival times of the shock at 1 AU as diamonds (stars for level 5 grid refinement and squares for level 2 grid refinement) in order for the 4 Sep 2017 CME (event 1), the 6 Sep 2017 CME, the 12 Feb 2018 CME, and the 29 Nov 2013 CME (event four). The observed and simulated arrival times for level 5 grid refinement match each other {\color{black} very well}. The predictions for level 2 grid refinement are less accurate because of much broader shock fronts.}
\end{figure}

{\color{black} Importantly, the increased grid resolution leads to a decreased initial CME acceleration with the CME launcher inbuilt in the code. This requires the subsurface current that generates the CME magnetic fields to be increased in order to obtain the observed initial CME outflow speed, as shown in Fig.~\ref{Figure_1}. 
Typical currents in our refined simulations are about $3 \times 10^{12}$ A, which are larger than the currents of $2.5 \times 10^{11}$ A - $6 \times 10^{11}$ A in \citet{Manchester08,Manchester12}.
\citet{Melrose17} states that the typical current for a flux rope in an active region driving a flare is $1 \times 10^{11}$ A, but he also states that this current can be much larger. \citet{Titov99} and \citet{Savcheva12} fitted observed flux ropes in an active region with a current of $7 \times 10^{12}$ A and $2 \times 10^{12}$ A, respectively. Thus, the electric currents in our refined simulations are in the observed range.
Interestingly, the currents in our level 2 runs are a factor of $\approx$ 10 smaller than for the level 5 runs (Fig.~\ref{Figure_1}). The values are in the range $1-4 \times 10^{11}$ A.  

{\color{black} A final refinement, of lesser importance here, is that before releasing the CME} we ran the code in time-independent mode for 1000 steps {\color{black} (rather than the default 300)} in order to obtain {\color{black} a more settled plasma-field system out to 1 AU.}  

\section{Simulation results}

The six CME events we simulated started on 4 Sep 2017 at 19:12 UT, 6 Sep 2017 at 12:30 UT, 7 Sep 2017 at 10:36 UT, 7 Sep 2017 at 15:24 UT, 12 Feb 2018 at 2:00 UT, and 29 Nov 2013 at 20:00 UT, based on CACTUS detections. The first five of these were launched close to the Sun-Earth direction, while the 29 Nov 2013 event headed toward STEREO A. In Fig.~\ref{Figure_2} we show colour-coded snapshots of the simulated magnetic field {\color{black} strength} in the ecliptic plane at a specific elapsed time {\color{black} for each} event. The yellow-bounded red features in each panel of Fig.~\ref{Figure_2} show the propagating CME and its driven shock. Clearly, the first five CMEs are predicted to hit the Earth and the sixth CME to hit STEREO A. 

Figures~\ref{Figure_3}(a) and (b) show the observed (solid lines) and simulated (stars) magnetic field strengths at the position of the Earth as functions of time for the two 7 Sep 2017 events. While the simulated CMEs do reach Earth, the simulated fields are always and usually well below the observed fields for the first and second events, respectively. This is consistent with the observations not {\color{black} showing} any signatures of a shock wave. The other MHD variables also show no observable shock or CME signatures. 

Fig.~\ref{Figure_4} presents the simulated and observed arrival times of the CME-driven shocks for the 4 Sep 2017, 6 Sep 2017, and 12 Feb 2018 CMEs at Earth (events 1 to 3) and at STEREO A for the 29 Nov 2013 CME (event four), defined as the temporal position halfway up the CME-driven shock ramp. In each case the difference between the simulated and observed shock arrival times is less than 2 hours. The {\color{black} average} of the absolute value of the difference is 0.9 $\pm$ 1.9 hours (mean absolute error plus or minus standard deviation) for these four events for level 5 initial  grid refinement simulations, and 1.9 $\pm$ 3.9 hours for level 2 initial grid refinement simulations. Thus, with the BATS-R-US code we predict the CME-driven shock arrival with a much better accuracy than the 10 $\pm$ 1 hours for ENLIL \citep[][]{Wold18}. 
{\color{black} We interpret the increased prediction accuracy of CME arrival times at 1 AU for our approach as a consequence of (a) demonstration of a linear relationship between the solar subsurface current and the average CME speed below 6 $R_s$, which allows us to fine-tune precisely the solar subsurface current to the CME outflow speed observed below 6 $R_s$, and (b) increasing the initial grid resolution by three levels, leading to sharper CME-driven shocks, less numerical dispersion (and viscosity), and changed propagation speeds.}
This type of linear relation likely exists for other simulation codes, suggesting that our approach for fine-tuning is likely widely applicable.}

\begin{figure}[!ht]
\includegraphics[scale=1.13]{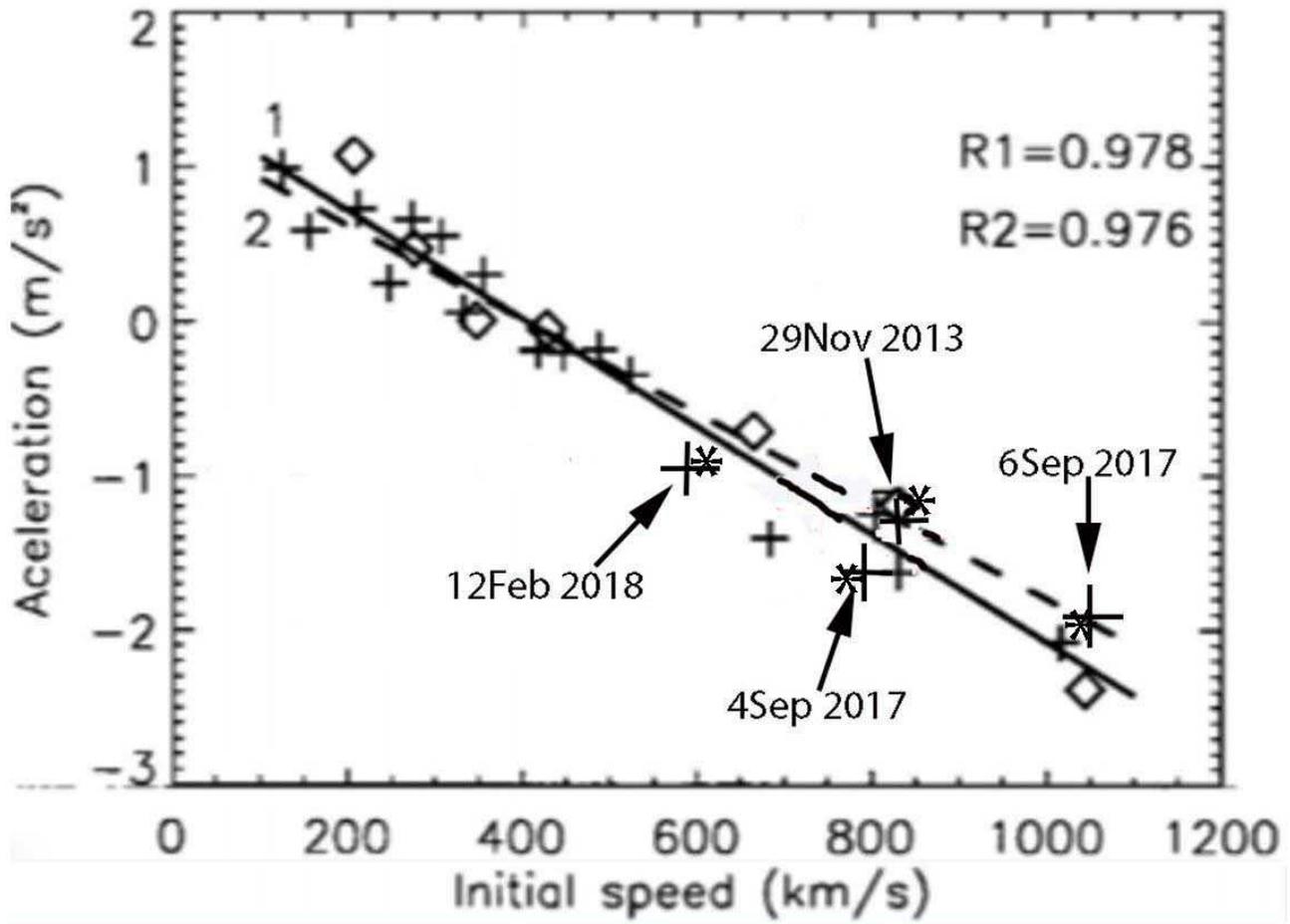}
\caption{\label{Figure_5} Simulation results for the initial shock speed and average acceleration for the 4 Sep 2017, 6 Sep 2017, 12 Feb 2018, and 29 Nov 2013 CME events, as derived from height - time measurements in the simulation box, are superposed onto Figure 2 of \citet{Gopalswamy00}. Crosses are for level 5 grid refinement simulations and stars for level 2 grid refinement simulations. The simulation results agree well with the empirical models of \citet{Gopalswamy00} (dashed and solid lines) and associated data (diamond and plus symbols).}
\end{figure}

Fig.~\ref{Figure_5} overplots the initial speeds and average accelerations for the CME events onto {\color{black} Figure 2} of \citet{Gopalswamy00}, which shows observational data and a fitted linear model between the observed initial CME speeds and average accelerations. We determined these parameters as the slope and curvature of CME height-time diagrams measured in the simulation box. We find that the simulation results are very close to the \citet{Gopalswamy00} model and associated data. Thus, the CMEs simulated with the BATS-R-US code {\color{black} appear to have very similar} dynamical properties {\color{black} to} observed CMEs.  

An issue with our level 5 simulations is that,
as a consequence of the large flux rope currents, the simulated CME magnetic fields near 1 AU are about ten times larger than the observed fields. Also, the present approach does not accurately predict the ambient solar wind plasma and field parameters in the ecliptic near 1 AU, as also found in many other papers for multiple simulation codes \citep[e.g.,][]{Cohen08,Jian11,Gressl12,Shen18,Den18,Torok18}. From an operational point of view such imprecisions in CME fields and solar wind properties are of lesser concern, compared with the gain in prediction accuracy of the CME arrival time at 1 AU demonstrated below for our simulations. Our opinion is that the greater priority for the space weather community is a tool to predict these arrival times more accurately, as described above. 

\begin{figure}[!ht]
\includegraphics[scale=0.18]{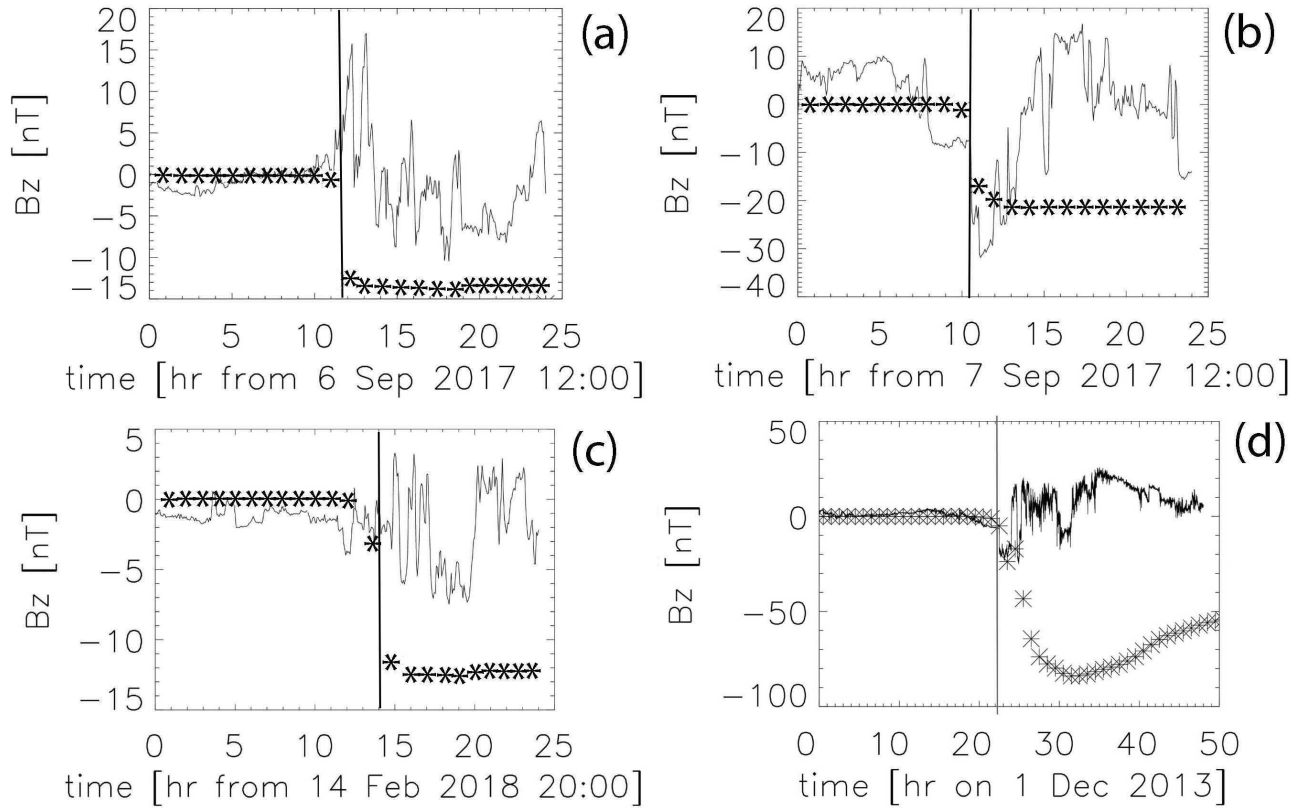}
\caption{\label{Figure_6} Simulated variations in $B_z$ (stars) at 1 AU for the (a) 4 Sep 2017, (b) 6 Sep 2017, (c) 12 Feb 2018, and (d) 29 Nov 2013 CME events. The solid lines are ACE (a)-(c) and STEREO A (d) {\color{black} $B_z$} measurements. Vertical lines denote the observed shock transition.}
\end{figure}

Turning now to the $B_z$ predictions, Fig.~\ref{Figure_6} shows the simulated (stars) and observed (solid line) $B_z$ magnetic field as a function of time at Earth for the 4 Sep 2017, 6 Sep 2017, and 12 Feb 2018 CME events and at STEREO A for the 29 Nov 2013 CME event. For the 4 Sep 2017 CME event in Fig.~\ref{Figure_6}(a) the simulated $B_z$ turns abruptly negative at the shock arrival time (hour 11.5 $\pm$ 0.5) and remains $<$ -15 nT until after hour 25. The observed $B_z$ component is highly disturbed just before and after the CME-driven shock: $B_z$ turns positive near hour 11, has two large oscillations, then turns strongly negative with large oscillations around hour 13.5, and remains predominantly negative until hour 23. The BATS-R-US simulation is not able to {\color{black} predict} the major oscillations of $B_z$ before or after the shock transition. However, the simulated change in $B_z$ onsets within 2 hours of the right time and is predominantly negative. Thus, ignoring the oscillations in $B_z$, the polarity of the change in $B_z$ appears to be predicted correctly, suggesting that this CME event should be geoeffective. An increase of the Kp index to 5 on 6 Sep 2017 8:00 UT was {\color{black} indeed} observed.

Similar patterns are found for the other events. 
For the 6 Sep 2017 CME event a strongly negative change in $B_z$ is predicted at the shock transition. This negative change coincides well with the observed change in $B_z$, which however shows major oscillations and becomes positive {\color{black} 3 hours later}. Thus, the onset of {\color{black} a} trigger {\color{black} for} space weather is predicted reasonably well for the 6 Sep 2017 event. 

The simulated $B_z$ component for the 12 Feb 2018 CME event also becomes strongly negative after the shock transition and remains there. The location of the large negative change in $B_z$ is very close to the onset of the observed CME-driven shock, after which the observed $B_z$ starts to oscillate {\color{black} substantially}. The observed oscillations in $B_z$ are not captured in the simulation.

Finally, the simulation for the 29 Nov 2013 CME event correctly predicts the onset and magnitude of a large negative change in $B_z$ shortly after the shock arrival, which is later completely obscured by very strong oscillations observed in $B_z$. If STEREO A were at Earth, it would be reasonable to predict a space weather event. 

Noting that resolving a change reliably requires at least 3 radial cells and so a convection time of $\approx 3 R_s / v_{\rm sw}$ $\approx$ 1.5 hours, it is not surprising that the code does not adequately resolve or simulate the large oscillations observed in $B_z$ on timescales of 1-2 hpurs. Moreover, given the large amplitude of these oscillations in the observed $B_z$ time series, it is unclear how well $B_z$ is modelled behind the shock. In detail, while the 6 Sep 2017 and 29 Nov 2013 events show the predicted change in $B_z$ directly behind the shock, the 4 Sep 2017 and 12 Feb 2018 events are obscured by large oscillations in $B_z$. Accordingly, no firm conclusion can be reached at this time whether the sign of $B_z$ is adequately predicted in our simulations.

\section{Discussion and conclusions}

We have simulated the propagation of six observed CMEs from {\color{black} the} Sun to either Earth or STEREO A with the 2012 BATS-R-US code and examined the implications for space weather {\color{black} predictions.} 
{\color{black} We developed a refined approach that determines the CME parameters using CAT and coronagraph data, and implements (and shows the necessity of) much higher initial grid refinement levels and numbers of simulation cells to reduce numerical dispersion and attain (relatively) sharp shocks, fine-tunes the CME launch using a new linear relationship between flux rope current and initial average CME speed below 6 $R_s$, and uses a more settled plasma-field system out to 1 AU. This approach for 6 events results in the greatly improved and accurate (error of $1 \pm 2$ hours) prediction of shock arrival times at 1 AU, accurately predicts whether or not a specific CME will hit an observer (Earth or STEREO A), and shows consistency of the simulated average speed and acceleration values with the model and data of \citet{Gopalswamy00}. It also shows promise in predicting the dominant sign of $B_z$ after shock arrival, although the trend in $B_z$ is obscured and made less reliable by large oscillations observed in the $B_z$ data. Therefore, our approach appears to} predict {\color{black} well} the onset of triggers {\color{black} for} possible space weather events. 

However, the BATS-R-US code does not simulate the large oscillations observed in $B_z$ in the solar wind or after the CME-driven shock arrival, {\color{black} although} for stronger CME events (the 6 Sep 2017 and 29 Nov 2013 events) the simulated and observed negative changes in $B_z$ are in better quantitative agreement. {\color{black} Of more general concern,} there {\color{black} are} major quantitative {\color{black} differences} between the simulated and observed magnetic fields {\color{black} and plasma parameters. These issues require further work.}     

Having steeper and more realistic CME-driven shocks in the simulation requires {\color{black} 5th level initial grid refinement and} 16 million simulation cells. In order to make the simulation operational, where the sum of the setup, simulation, and analysis times is much less than the propagation time of the CME from the Sun to 1 AU, requires {\color{black} use} of a large parallel computer. For our cases, {\color{black} with 96 processors,} these times were $\approx$ 1 hour, $\approx$ 6 hours, and $\approx$ 5 hours, so that operational use appears possible. 

In conclusion, since the CME speeds, accelerations and arrival times are predicted {\color{black} well} with the 2012 BATS-R-US code {\color{black} using our refined approach}, with some promise also for the qualitative changes in $B_z$ upon arrival, it appears that {\color{black} this} code {\color{black} and refined setup approach} are appropriate for space weather predictions and should be explored further.


%
%
%
%
%
%
%

\begin{acknowledgments}
We thank G. T\^{o}th, {\color{black} B. van der Holst}, T. Gombosi, and other Michigan colleagues for their assistance and for providing the {\color{black} 2012} BATS-R-US code, {\color{black} and N. Gopalswamy for advice.} This work was funded by Australian Research Council grant DP180103509 and the Space Weather Services unit of Australia's Bureau of Meteorology, with NCI providing HPC resources. Funding agencies for the Space Weather Modeling Framework (Center for Space Environment Modeling, University of Michigan) are NASA ESS, NASA ESTO-CT, NSF KDI, and DoD MURI. CME data were taken from the archives ftp://zeus.nascom.nasa.gov/qkl/lasco/quicklook/level\_05/ and https://www.nasa.gov/mission\_pages/stereo/multimedia\/image-archive.html.
\end{acknowledgments}

\end{article}
%
%
%
%
%
%
%

\newpage

\end{document}